\documentclass[a4paper,12pt]{article}

\usepackage[utf8]{inputenc}
\usepackage[english]{babel}
\usepackage[pdftex,colorlinks,linkcolor=blue,citecolor=blue]{hyperref}
\usepackage[pdftex]{graphicx}

\usepackage[T2A]{fontenc}
\usepackage{amsmath,amssymb,amsfonts,amsthm}
\usepackage{pgfplots}
\usepackage{tikz}
\usepackage[affil-it]{authblk}

\title{Charged topological geons with\\ \vspace{-0.5ex}a self-gravitating scalar field}
\author{\vspace{-1ex}Alexander Tsirulev\thanks{tsirulev.an@tversu.ru}}
\affil{\vspace{-1ex}\small Faculty of Mathematics, Tver State University, Tver, Russia, 170002}
\date{}

\begin{document}

\maketitle
\renewcommand{\abstractname}{}\abstractname

\vspace{-12ex}
\begin{abstract}\noindent
A topological geon is an asymptotically (anti-)de Sitter or flat spacetime with topology $\mathbb{R}\times{}M$, where $M$ is the punctured projective space, ${M=\mathbb{R}P^3\backslash\{p\}}$, and the removed point $p$ corresponds to the spacelike infinity. The manifold $M$ is conventionally obtained from a spherically symmetric spatial slice of a wormhole spacetime as the quotient manifold by the isometric action of the group $\mathbb{Z}_2$. We study the general properties of static, spherically symmetric, charged topological geons supported by a selfgravitating, minimally coupled scalar field with negative kinetic energy and an arbitrary selfinteraction potential. In the most general case, it turns out that the gravitational mass of such a geon is completely determined by its electric charge and size (for a given scalar field), that is, the size of the corresponding wormhole throat. We discuss the properties of the charged Ellis-Bronnikov-Sorkin geon and argue that it can be considered as a possible classical model of elementary particles beyond the Standard Model, in particular, as dark matter particles.
\end{abstract}

{\small {\bf Keywords:} topological geon, wormhole, scalar field, Einstein-Maxwell-Klein-Gordon equations.}

\section{Introduction}

In gravitational physics, the assumption about the possibility of nontrivial spacetime topology has a long history~\cite{Einstein1935, Fuller1962}, but nowadays researchers working in this direction are mainly interested in wormholes~\cite{Thorne1988, Visser1995}, which connect two spatially separated regions in the same spacetime. Traversable wormholes are known to be supported by some exotic matter that violates the null energy condition, in particular, by a phantom scalar field acting as a source~\cite{Ellis1973, Bronnikov1973, Bronnikov2023-1}. Phantom scalar fields have the opposite---in comparison with classical scalar fields---sign of the kinetic term in its Lagrangian and arise naturally in modern cosmological models. So far we have no evidence for the existence of wormholes, for example, in the centers of galaxies, but various theoretical models predict many observational effects that would distinguish wormholes from other astrophysical objects~\cite{Novikov2010, Sushkov2023, Bronnikov2023-2, Bronnikov2025, Potashov2020}\;\!\footnote{This paper is not a review, therefore it is impossible to cite the dozens of meaningful works on this issue: the bibliography reflects the author's personal preferences.}. Geometrically, a wormhole cannot be considered as a particle in space-time, although the Einstein-Rosen wormhole~\cite{Einstein1935} with a singularity at its throat was constructed as a model of a particle. In contrast, a topological geon, introduced in gravitational physics by R.~D.~Sorkin~\cite{Sorkin1986}, possesses the main features of a particle at least on a theoretical level.

In this paper, we study a model of a static, spherically symmetric, charged topological geon supported by a selfgravitating phantom scalar field with an arbitrary self-interaction potential. One of our motivations for studying such exotic configurations is based on the intuitive belief that simple solutions to fundamental equations are very likely to be realized in nature. Another motivation is related to the problem of dark matter, whose particles appear to be beyond the Standard Model. We do not consider the processes of geon formation, since they are inevitably speculative. However, if the quantum state is assumed to be a fundamental and universal entity, then the concept of quantum transitions can also be applied to various spacetime topologies, so that topological transitions could have occurred in the very early universe.

The paper is organized as follows. In Sec.~\ref{sec2}, we consider in detail the topological structure of spherically symmetric wormholes and geons with two- and three-dimensional spatial slices, as well as the one-to-one correspondence between geons with topology ${\mathbb{R}\times\mathbb{R}P^3\backslash\{p\}}$ and wormholes with topology ${\mathbb{R}^2\times\mathbb{S}^2}$ and central symmetry. In Sec.~\ref{sec3}, an analytical model for such wormholes and the corresponding geons is constructed in the form of quadratures for an arbitrary scalar field potential. It is shown that these quadratures establish a connection between the mass, charge and scalar field parameters of a topological geon. In Sec.~\ref{sec4} we discuss an important example of a charged geon, which is called below the Ellis-Bronnikov-Sorkin geon for the following reason: the corresponding wormhole is a generalization of the Ellis-Bronnikov wormhole to its charged version. We also discuss the main features of this geon in the context of the classical modelling of elementary particles. Sec.~\ref{sec5} provides concluding remarks and a brief discussion of the results obtained in this work.

Throughout this paper, we adopt the signature $(+,-,-,-)$ for the metric and use the geometrized units in which $G=c=1$. Latin indices run from 0 to 3 and are used to denote spacetime components of tensors, while Greek indices take the values 1,2,3 and denote purely spatial components.

\section{Topological structure of wormholes and geons}
\label{sec2}

The concept of a topological geon was introduced in~\cite{Sorkin1986} as a spacetime with topology ${\mathbb{R}\times\mathcal{M}}$, where the quotient $\mathcal{M}$ is a punctured compact three-dimensional manifold with nontrivial topology (that is, $\mathcal{M}$ is not ${\mathbb{S}^3\backslash\{p\}}$), and a removed point corresponds, from a geometric point of view, to a flat or (anti-)de Sitter asymptotic infinity. Usually this notion implicitly assumes that the topologically nontrivial part of a geon is spatially bounded and can be enclosed in a ball of finite size, so that geons can be thought of as particles consisting of spatial topology~\cite{Sorkin1998}.

Topological geons of this general form were considered in~\cite{Louko2005}, but an analytical study is possible so far only for the ${\mathbb{R} \times \mathbb{R}P^3\backslash\{p\}}$ geon; this paper is devoted specifically to this configuration without an event horizon. It turns out that such geons are in one-to-one correspondence with traversable wormholes that possess topology ${\mathbb{R} \times \mathbb{R}^3\#\mathbb{R}^3 \approx \mathbb{R}^2\times\mathbb{S}^2}$ and the isometry group $\mathbb{Z}_2$ acting in spherical coordinates by the transformation
\begin{equation}\label{}
\big(t,r,\theta,\phi\big) \rightarrow \big(t,-r,\pi-\theta,\phi+\pi\big).
\nonumber
\end{equation}
\begin{center}
\begin{figure}
\includegraphics[width=15.5cm]{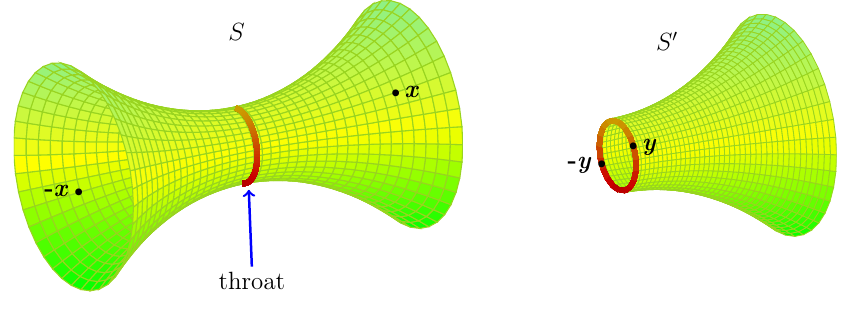}
\caption{This illustration shows the two-dimensional wormhole-geon correspondence. Here $S'$ is the $\mathbb{Z}_2$-quotient of $S\backslash\{\mathrm{troat}\}$. The spatial slice of the corresponding geon is obtained by identifying opposite points on the boundary circle in $S'$. Wormholes have topology ${\mathbb{R}\times\mathbb{R}^2\#\mathbb{R}^2}$, while the resulting geons are homeomorphic to ${\mathbb{R}\times\mathbb{R}P^2\backslash\{p\}}$.}
\label{fig1}
\end{figure}
\end{center}

\vspace{-2ex}
For a (1+2)-dimensional gravity, a spatial slice of such a wormhole has topology ${\mathbb{R}^2\times\mathbb{S}^1}$ and so can be embedded into ${\mathbb{R}^3}$, as is illustrated in Fig.\;\ref{fig1}. This figure also shows that the spatial slice of the corresponding (1+2)-dimensional topological geon is homeomorphic to the ${\mathbb{R}P^2\backslash\{p\}}$ and can be obtain by cutting an open disk out of the plane ${\mathbb{R}^2}$ followed by identifying the opposite points of the boundary circle.

\begin{center}
\begin{figure}[!h]
\includegraphics[width=0.85\textwidth]{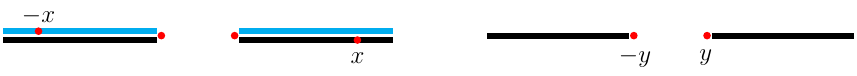}
\caption{These diagrams illustrate the three-dimensional wormhole-geon correspondence as it looks like for a single line passing through the center.}
\label{fig2}
\end{figure}
\end{center}

Analogously, a spatial slice (with topology ${\mathbb{R}P^3\backslash\{p\}}$) of a (1+3)-dimensional geon can be thought of as constructed from ${\mathbb{R}^3}$ by removing an open ball and subsequently identifying the opposite points of the boundary sphere. This procedure is schematically sketched in Fig.\;\ref{fig2}, where a wormhole is shown as it would look if it were submersed (as a set, not as a topological space) in ${\mathbb{R}^3}$. The spatial slice of a wormhole, ${\mathbb{R}^3\#\mathbb{R}^3}$, is represented in $\mathbb{R}^3$ as follows: in ${\mathbb{R}^3\backslash\mathbb{B}^3}$, all black points are doubled, except for points on the boundary of the open ball $\mathbb{B}^3$. The intermediate quotient is obtained by identifying each point $x$ (from the initial set ${\mathbb{R}^3\backslash\mathbb{\overline{B}}^3}$) with its opposite point $-x$ (from the set of copied points). Finally, in the right diagram, the corresponding geon is constructed by identifying opposite points on the two-dimensional boundary sphere in ${\mathbb{R}^3\backslash\mathbb{B}^3}$.

A geon with topology ${\mathbb{R}\times\mathbb{R}P^3\backslash\{p\}}$ is time and space orientable; the latter follows from the fact that the manifold $\mathbb{R}P^3$ is diffeomorphic to the Lie group ${SO(3,\mathbb{R})}$. However, the point ${r=0}$ in spherical coordinates corresponds to the non-orientable surface $\mathbb{R}P^2$, as can be seen from Fig.\;\ref{fig2}. This implies that in the spacetime of a spherically symmetric topological geon, the electromagnetic field, given by a two-form ${F=\alpha{dt}\wedge{dr}}$, takes its values in a vector bundle with the structure group $\mathbb{Z}_2$. The same is also true for scalar fields with odd field functions, ${\phi(-r)=-\phi(r)}$, while even field functions are sections in the trivial vector bundle with fiber $\mathbb{R}$. These features of the fields become obvious if one considers these fields in the standard atlas consisting of six charts on ${\mathbb{R}P^3\backslash\{p\}}$ or, equivalently, on $\mathbb{R}P^3$ (a minimal atlas on $\mathbb{R}P^3$ consists of four charts).

\section{General analytical model}
\label{sec3}

We consider the simplest model with the action
\begin{equation}\label{action}
    \Sigma= \int\left(-\frac{1}{2}S- \langle d\phi,d\phi\rangle-2V(\phi)- \frac{1}{2}\langle \mathcal{F},\mathcal{F}\rangle\,\right) \sqrt[]{|g|}\,d^{\,4}x\,,
\end{equation}
where $S$ is the scalar curvature, $V(\phi)$ is a potential representing the selfinteraction of the scalar field $\phi$, $\mathcal{F}$ is an electromagnetic 2-form, and angle brackets denote the scalar product induced by the spacetime  metric $g$. Note first that the scalar field kinetic term is negative, and second, that the cosmological constant $\Lambda$---if considered to be nonzero---is included in the potential.

We consider spherically symmetric, static, \textit{traversable} wormholes that have reflection symmetry with respect to the centers of their throats, as well as the corresponding topological geons. Therefore, it is convenient to use the so-called quasiglobal coordinates, in which the metric has the form
\begin{equation}\label{metric}
ds^2=Adt^2-\frac{\,dr^2}{\,A}-C^{2} \left(d\theta^2+\sin^{2}\theta\,d\varphi^2\right),
\end{equation}
where the \textit{positive} metric functions $A$ and $C$, and the field $\phi$, depend only on the radial coordinate $r$ in the range from $-\infty$ to $\infty$. It is natural to introduce the orthonormal bases associated with the metric~(\ref{metric}), namely, the basis of vector fields,
\begin{equation}\label{}
\left\{\mathrm{e}_0= \frac{1}{\sqrt{A}}\,\partial_{t},\,\; \mathrm{e}_1= \sqrt{A}\,\partial_{r},\,\;
\mathrm{e}_2= \frac{1}{C}\,\partial_{\theta},\,\; \mathrm{e}_3=
\frac{1}{C\sin\theta}\,\partial_{\varphi}\right\},
\nonumber
\end{equation}
the dual basis $\big\{\mathrm{e}^0 = \sqrt{A}\:\!dt,\ldots\big\}$ of 1-form, and the corresponding basis of 2-form,
\begin{equation}\label{}
\left\{\alpha^{1}= \mathrm{e}^{0}\!\wedge\;\!\!\mathrm{e}^{1},\;\;
\alpha^{2}= \mathrm{e}^{0}\!\wedge\;\!\!\mathrm{e}^{2},\,\,
\alpha^{3}= \mathrm{e}^{0}\!\wedge\;\!\!\mathrm{e}^{3},\;\;
*\,\alpha^{1}= \mathrm{e}^{3}\!\wedge\;\!\!\mathrm{e}^{2},\;\;
*\,\alpha^{2}= \mathrm{e}^{1}\!\wedge\;\!\!\mathrm{e}^{3},\;\;
*\,\alpha^{3}= \mathrm{e}^{2}\!\wedge\;\!\!\mathrm{e}^{1}\right\},
\nonumber
\end{equation}
where $*$ is the Hodge star operator. Further, we find nonzero connection forms, defined by ${\nabla^{\vphantom{\int}}_{\!\!\vphantom{\int}\scriptstyle
X}e_{j}=\omega^i_j(X)\,\mathrm{e}_{i}\!\:}$, where $\omega_{0}^{\alpha}=\omega_{\alpha}^{0}$ and $\omega_{\beta}^{\alpha}=-\omega_{\alpha}^{\beta}$ for any orthonormal basis of vector fields. Then we can find the Riemann curvature\footnote{In this paper, the signs and positions of the indices in the curvature and Ricci curvature are fixed in the usual manner: $R(X,Y)=\nabla_X\nabla_Y-\ldots\,$, $R(\omega,Z,X,Y)=\langle\omega,R(X,Y)Z\rangle$, and $\mathcal{R}_{jl}=R^i_{jil}$.} using the first and second Cartan equations of structure (see, e.g., \cite{Chandra2001}). We obtain the expressions
\begin{equation*}
\omega_{0}^{1}=\frac{A'}{2\sqrt{A}}\,\mathrm{e}^0,\quad
\omega_{1}^{2}=\frac{\sqrt{A}C'}{C}\,\mathrm{e}^2,\quad
\omega_{1}^{3}=\frac{\sqrt{A}C'}{C}\,\mathrm{e}^3,\quad
\omega_{2}^{3}=\frac{\cot\theta}{C}\,\mathrm{e}^3,
\nonumber
\end{equation*}
\begin{multline}\label{R}
R=-\frac{A''}{2}\,\alpha^1\!\otimes\alpha^1- \frac{A'C'}{2C}
\left(\alpha^{2}\otimes\alpha^{2}+ \alpha^{3}\otimes\alpha^{3}\right)\\
+\frac{A{C'}^2-1}{C^2}*\!\alpha^1\!\otimes*\alpha^1+ \frac{2AC''+A'C'}{2C}
\left(*\alpha^{2}\otimes*\alpha^{2}+*\alpha^{3}\otimes*\alpha^{3}\right),
\end{multline}
where a prime denotes differentiation with respect to $r$. From the latter expression, it is easily to compute the left-hand side of the Einstein equations ${\mathcal{R}-\big(S/2\big)g=\mathcal{T}^{(\phi)}+\mathcal{T}^{(e)}}$.

The Klein-Gordon equation (with a negative kinetic term) reads $\Box\phi-dV/d\phi=0$. For static spherically symmetric configurations, the energy-momentum tensor of the scalar field has the form
\begin{equation}\label{F}
\Big(\mathcal{T}_{ij}^{(\phi)}\Big)=  \mathrm{diag}\big(-A{\phi'}^2+ 2V,\; -A{\phi'}^2- 2V,\; A{\phi'}^2- 2V,\; A{\phi'}^2- 2V\big).
\nonumber
\end{equation}
Electromagnetic potentials compatible with spherical symmetry must necessarily have the form ${\mathcal{A}=\sigma(r\;\!\!)\:\!dt}$, and the electromagnetic 2-form is defined by the equations ${\mathcal{F}=d\mathcal{A},\; d*\mathcal{F}=0}$; their solution and the corresponding energy-momentum tensor are given by the expressions
\begin{equation}\label{F}
\mathcal{F}= \frac{q}{C^{2}}\,\mathrm{e}^0\!\wedge\;\!\!\mathrm{e}^1, \qquad \Big(\mathcal{T}_{ij}^{(e)}\Big)=  \mathrm{diag}\big(q^2/C^4,\; -q^2/C^4,\; q^2/C^4,\; q^2/C^4\big).
\nonumber
\end{equation}

Now the independent Einstein-Klein-Gordon-Maxwell field equations for the action~(\ref{action}) can be explicitly written as
\begin{eqnarray}\label{00}
-2A\frac{\,C''}{C}- A'\,\frac{\,C'}{C}- A\frac{{\,C'}^2}{C^2}+ \frac{1}{\,C^2} &=& -A{\phi'}^2+ 2V+ \frac{q^2}{C^4}, \\
\label{11}
A'\frac{\,C'}{C} + A\frac{{\,C'}^2}{C^2} - \frac{1}{\,C^2} &=& -A{\phi'}^2 -2V- \frac{q^2}{C^4}, \\
\label{KG}
A\phi''+ \phi'\!\left(A'+ 2A\frac{\,C'}{C}\right)+ \frac{dV}{d\phi} &=& 0.
\end{eqnarray}
Without loss of generality, one can adopt $r=0$ at the throat of a wormhole (which is assumed to be symmetric with respect to its throat), so that $A$ and $C$ are even functions, while $\phi$ can be either even or odd. Under the assumption of asymptotic flatness of spacetime, the natural conditions at spatial infinity are
\begin{equation}\label{asympt}
\phi= O\!\left(1\right), \quad A=1-\frac{2M}{|r|}+O\!\left(|r|^{-2}\right), \quad C=|r|+b+O\!\left(|r|^{-1}\right),\quad r\rightarrow\pm\infty,
\end{equation}
where $M$ is the gravitational mass as measured by a distant observer.

The form of the potential $V(\phi)$ is unknown a priori. Therefore, we use the reconstructed potential method (also referred to as "inverse problem method for self-gravitating scalar fields") to model wormhole and geon configurations, in some sense, simultaneously for all physically reasonable self-interaction potentials. In implicit form, this method was proposed in~\cite{Lechtenfeld1998, BronnikovShikin2002, Nikonov2008, Tchemarina2009, Azreg2010}, and then obtained in~\cite{Solovyev2012} in the form of explicit quadratures; for the problem (\ref{00})\,--\,(\ref{asympt}) with the metric~(\ref{metric}), they are given by the expressions
\begin{equation}\label{q1}
\phi'= \sqrt{C''/C\,},
\end{equation}
\begin{equation}\label{q2}
Q(r)= 2q^2\!\!\int\limits_{\!r}^{\,\,\infty}
\!\frac{dr}{C^2},
\end{equation}
\begin{equation}\label{q3}
A=2C^2\!\!\int\limits_{\!r}^{\,\,\infty}\! \frac{\,r+Q-a}{\,C^4}\,dr-
\frac{\,\Lambda C^2}{3},
\end{equation}
\begin{equation}\label{q4}
\widetilde{V}(r)= \frac{1}{2C^2}
\left(1-3{C'}^2A- CC''A+ 2C'\,\frac{Q-a}{C}- \frac{q^2}{C^2}\right),
\end{equation}
where ${a\in\mathbb{R}}$. It can be verified by direct differentiation that these quadratures reduce equations (\ref{00})\,--\,(\ref{KG}) to identities. In (\ref{q3}) and (\ref{q4}), for generality, the cosmological constant, for a moment, is taken into account and incorporated in the potential such that ${V\big(\phi(\infty)\big)=\Lambda/2}$. In what follows, for definiteness, we restrict our attention to asymptotically flat wormholes and geons, since their "intrinsic properties"\;coincide with the properties of the corresponding (anti-)de Sitter configurations, as can be seen from the expression~(\ref{q3}).

In the context of models of wormholes, the quadratures (\ref{q1})\,--\,(\ref{q4}) should be used as follows: as a first step, a positive function $C(r)$ of class $\mathcal{C}^2$ is chosen such that (i)~$C(r)=C(-r)$, (ii)~$C(r)$ and $C'(r)$ monotonically increase in the interval ${(0,\infty)}$, and (iii)~the last condition in~(\ref{asympt}) holds; in the second step, the functions $\phi$, $Q$, and $A$ should be sequentially calculate; and finally, inverting the monotone function ${\phi(r)}$, we obtain the reconstructed potential as ${V(\phi)=\widetilde{V}(r(\phi))}$. It is straightforward to show that the gravitational mass in asymptotically Schwarzschild-like coordinates (in which $C=|r|$ outside the central region) can be expressed in terms of the constants $a$ and $b$ as
\begin{equation}\label{m}
M= \frac{a+b}{3}\,.\nonumber
\end{equation}
When an even function $C(r)$ is given, the requirement for the wormhole to be symmetric with respect to the center of its throat is equivalent to the requirement that the integrand in~(\ref{q3}) be an odd function. This, in turn, implies $a=Q(0)$, and consequently
\begin{equation}\label{m-q}
M= \frac{Q(0)+b}{3}= \frac{2q^2}{3} \!\!\int\limits_{\!0}^{\,\,\infty} \!\frac{dr}{C^2} + \frac{b}{3}\,.
\end{equation}
This equality is a necessary condition for the existence of charged topological geons with the simplest topology ${\mathbb{R}\times\mathbb{R}P^3\backslash\{p\}}$.

Thus, the main features of the model under consideration can be briefly expressed as follows: \textit{the mass of a topological geon is related to its charge and scalar field by means of the integral formula}~(\ref{m-q}) \textit{in which the term $b/3$ represents "the geometric part" of the mass, similar to the mass in the vacuum Schwarzschild solution; on the other hand, the scalar-electromagnetic part of the mass, $Q(0)/3$, is ultimately determined by the distribution of the gradient of the scalar field} (\textit{via equation}~(\ref{q1})), \textit{and electromagnetism is included in it simply as a weighting factor, namely, as the square of a charge}.

\section{Ellis-Bronnikov-Sorkin geon}
\label{sec4}

The simplest example of a charged wormhole and the corresponding topological geon is provided by the ansatz
\begin{equation}\label{C-EBq}
C(r)=\sqrt{r^2+\delta^2},
\end{equation}
where ${\delta=C(0)}$ is the size of the wormhole throat and also the size of the geon. This choice of the function $C(r)$ allows us to treat the quadratures (\ref{q1})\,--\,(\ref{q4}) analytically. We obtain
\begin{equation}\label{M-EBq}
M=\frac{\pi{}q^2}{3\delta},
\end{equation}
\begin{equation}\label{A-EBq}
A=1+\frac{q^2}{\delta^2}\left\{\left[1+ \frac{r}{\delta}\, \arctan\! \left(\frac{r}{\delta}\right)\right]^{2}+ \arctan^2\! \left(\frac{r}{\delta}\right) - \frac{\pi^2}{4}\! \left(\!1+ \frac{r^2}{\delta^2}\right)\!\right\}.
\end{equation}
This solution is obviously obeys the condition of the reflection symmetry $A(-r)=A(r)$. In the asymptotically flat region, $A(r)$ close to the Reissner-Nordstr\"{o}m solution with the mass~(\ref{M-EBq}), that is, $b=0$ and
\begin{equation}\label{A-asympt}
A=1-\frac{2\pi{}q^2}{3\delta r}+ \frac{q^2}{r^2}+ O\big(r^{-3}\big),\;\; r\rightarrow\infty.
\end{equation}

Near the center, the series expansion of $A(r)$ to second order is
\begin{equation}\label{A-asympt-r}
A= 1+ \frac{q^2}{\delta^2}\left(1- \frac{\pi^2}{4}+ \left(3- \frac{\pi^2}{4}\right)\frac{r^2}{\delta^2}\right)+ O\big(r^{3}\big),\;\; r\rightarrow0.
\end{equation}
From here, we obtain an additional restriction on the magnitudes of $q$ and $M$ for the solution~(\ref{A-EBq}): the condition $A(r)>0$ requires that
\begin{equation}\label{A>0}
q< \delta \left(\frac{\pi^2}{4}-1\right)^{\!-1/2}\! \approx0.826\;\!\delta, \qquad M< \delta\,\frac{\pi}{3}\! \left(\frac{\pi^2}{4}-1\right)^{\!-1} \approx0.714\;\!\delta.
\end{equation}
If this inequalities do not hold, then the solution~(\ref{A-EBq}) represents a regular black hole with the same nontrivial topology (hidden inside the event horizon), and the corresponding wormhole is not traversable. Note also that the metric function $A(r,\delta)$ given by the expression~(\ref{A-EBq}) remains regular as ${\delta\rightarrow0}$ (and, consequently, ${q\rightarrow0}$) in the sense that it is continuous on the interval ${0<r<\infty}$ and ${A=1,\, M=0}$ in the limit. The latter follows directly from~(\ref{A-EBq}) or the power series expansion
\begin{equation*}
A= -\frac{2q^2\pi}{3r}\,\delta^{-1}+ 1+ \frac{q^2}{r^2}+ \frac{2q^2\pi}{15r^3}\,\delta+ O\!\left(\delta^2\right),\quad \delta\rightarrow0,
\end{equation*}
if we take into account the first inequality in~(\ref{A>0}).

To complete the program of constructing solutions using quadratures (\ref{q1})\,--\,(\ref{q4}) as described above, it is necessary to find the scalar field function $\phi(r)$ and the selfinteraction potential $V(\phi)$. From~(\ref{q1}) and~(\ref{C-EBq}), it is easy to compute
\begin{equation}\label{phi}
\phi=\arctan\frac{r}{\delta}.
\end{equation}
Thus, $\phi(\pm\infty)=\pm\pi/2$ and $\phi(0)=0$. The function $\phi(r)$ is odd and monotone, and takes its values in a one-dimensional real vector bundle with the structure group $\mathbb{Z}_2$. Then the potential can be found by calculating $\widetilde{V}(r)$ followed by substituting the inverse function $r=\delta\tan(\phi)$ into the result. We obtain
\begin{multline}\label{V}
V(\phi)= \frac{\cos^4\!\phi- \cos^2\!\phi}{\delta^2}\\
+\frac{q^2}{\delta^4}\!\left(\frac{\cos^2\!\phi}{2}+ \left(\phi^2-\frac{\pi^2}{4}- \frac{3}{2}\right)\cos^2\!\phi- 3\phi\cos\!\phi\sin\!\phi- \frac{3\phi^2}{2}+ \frac{3\pi^2}{8}\right).
\end{multline}
It is easy to see that at spatial infinity ${V\big(\phi(\infty)\big)= V(\pi/2)=0}$. For $q=0$, we have the Ellis-Bronnikov massless wormhole and the corresponding uncharged topological geon with $A=1$.

It is well known that neither the Reissner-Nordstr\"{o}m nor the Kerr-Newman black holes cannot be considered as a model of an electron or any other charged elementary particle, since this is possible only if $q\ll M$, that is, for extremely large values of mass in comparison with its charge. However, for the Ellis-Bronnikov-Sorkin topological geon with the electron charge magnitude ${q_e\approx1.3\times10^{-34}\:\!\mathrm{cm}}$ and mass ${M_e\approx6.7\times10^{-54}\:\!\mathrm{cm}}$ (in the geometrized units), we find that the geon size is of quite reasonable value
\begin{equation}\label{delta-e}
\delta_e\approx2.6\times10^{-15}\:\!\mathrm{cm}.
\end{equation}
Note that the order-of-magnitude estimate is the same also for the $u$-quark.

If the topological geon is viewed as a toy model for dark matter particles, then it is more interesting to consider the family of solutions with a critical ratio of its mass, charge, and size, which corresponds to the signs of equality in~(\ref{A>0}). With these critical values, taking the mass M as the basic parameter, we obtain a \textit{geon of minimal size}, while its charge is then determined from the expression~(\ref{M-EBq}):
\begin{equation}\label{M-q-delta}
\delta= \frac{3}{\pi}\! \left(\frac{\pi^2}{4}-1\right)\!M \approx1.40\:\!M,
\qquad
q= \frac{3}{\pi}\! \left(\frac{\pi^2}{4}-1\right)^{\!1/2}\!M\approx1.16\:\!M.
\end{equation}

\begin{center}
\begin{figure}[!h]
\includegraphics [width=0.47\textwidth]{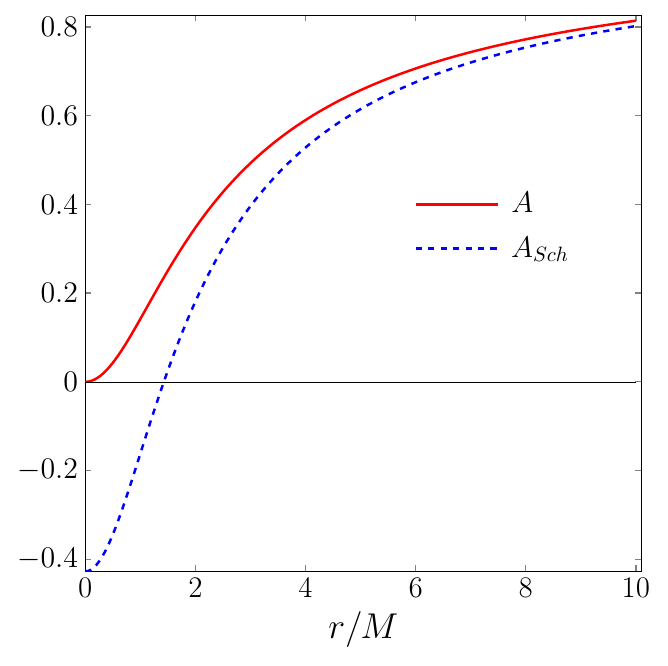}\;
\includegraphics [width=0.51\textwidth]{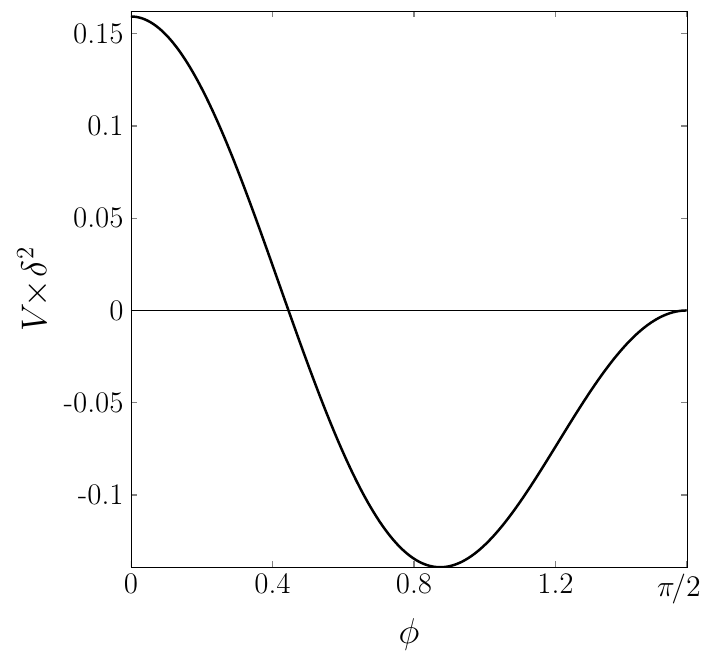}
\caption{Left panel: The metric functions $A(r)$ (for a geon with critical parameters---solid curve) and $A_{Sch}$ (for a Schwarzschild black hole with the same gravitational mass---dashed curve) expressed in natural wormhole coordinates, in which the area metric function has the form~(\ref{C-EBq}); the values of the radial coordinate are given in units of $M$. Right panel: The potential $V$ (given by the expression~(\ref{V}) for a geon with critical parameters) measured in units of $\delta^{-2}$.}
\label{fig3}
\end{figure}
\end{center}

Substituting~(\ref{M-q-delta}) into the expression~(\ref{A-EBq}) yields $A(0)=0$, so that the projective plane $\mathbb{R}P^2$ at $r=0$ behaves like an event horizon. Indeed, the Kretschmann invariant $\mathcal{K}$ near the center of the geon can be easily calculated using the formula~(\ref{R}); due to the cumbersomeness of the exact expression, we present only the first two terms of the Taylor series:
\begin{multline}\label{}
\mathcal{K}=R\cdot\!R= \left(\frac{3\pi^4}{4}-10\pi^2+44\right)\!\frac{q^4}{\delta^4}- \big(4\pi^2-16\big)\:\!\frac{q^2}{\delta^2}+ 12\\ +\left(\big(2\pi^2-8\big)\frac{q^2}{\delta^2}-8\right)\! \frac{r}{\delta}+O(r^2),\quad r\rightarrow0.
\end{multline}
Consequently, this invariant is regular everywhere, so that the singularity of the function $1/A$ at ${r=0}$ is purely coordinate. In order to obtain a geon that is close to critical and has a traversable center, it is sufficient to change $\delta$ to $\delta+\varepsilon$, where $\varepsilon\ll\delta$.

In particular, if $M$ is the Planck mass $M_{Pl}$, the geon size and charge are very close to the Planck length $L_{Pl}$ and the Planck charge $Q_{Pl}$, respectively. All these Planck constants, when expressed in the geometrized units, have the same magnitude, namely
\begin{equation*}
M_{Pl}=L_{Pl}=Q_{Pl}=\sqrt{\hbar} \approx 1.62\times10^{-33} \:\!\mathrm{cm},
\end{equation*}
so that the size and charge of the Ellis-Bronnikov-Sorkin geon are, respectively,
\begin{equation}\label{h-e}
\delta_{Pl} \approx 2.26\times10^{-33}\:\!\mathrm{cm}\:\!,
\quad
q_{Pl}\approx 1.87\times10^{-33}\:\!\mathrm{cm}\:\!.
\end{equation}

In the natural wormhole coordinates consistent with the metric~(\ref{metric}), the Schwarzschild metric has the form
\begin{equation*}
ds^2= A_{Sch}dt^2- \frac{1}{A_{Sch}}\:\!\frac{r^2}{r^2+\delta^2}\:\!dr^2- \big(r^2+\delta^2\big)d\Omega^{\:\!2},\quad A_{Sch}= 1-\frac{2M}{\sqrt{r^2+\delta^2}}.
\end{equation*}
For the critical parameters $M$, $q$, and $\delta$, the behavior of the geon metric function $A$ in comparison with $A_{Sch}$, as well as the shape of the selfinteraction potential, are plotted in Fig.\;\ref{fig3}. It is easy to see that $A$ and $A_{Sch}$ are very close already at $r\sim10M$. It should also be noted that the Reissner-Nordstr\"{o}m solution with such values of mass and charge is a naked singularity.

\section{Conclusions}
\label{sec5}

In this paper, we analytically study a two-parameter family of static, spherically symmetric, and everywhere-regular geons with the topology ${\mathbb{R} \times \mathbb{R}P^3\backslash\{p\}}$, supported by an electric field and a self-gravitating, minimally coupled phantom scalar field with an arbitrary self-interaction potential. We provide a complete characterization of such configurations: in particular, they exhibit particle-like properties and describe an asymptotically flat spacetime in which the electric field at spatial infinity looks like the field of a point charge. Using the method of reconstructing the scalar field potential, we obtain a general solution for wormholes and the corresponding geons in the form of quadratures, which allows us to obtain any specific solution by appropriately choosing the area metric function. We show that the gravitational mass of a topological geon supported by a given scalar field is completely determined by its electric charge and size.

In this work, we also study in detail the properties of the Ellis-Bronnikov-Sorkin geon and argue that it can be considered as a possible classical model of elementary particles beyond the Standard Model, in particular, as dark matter particles. In this connection, two comments should be made about this model. First, we have not discussed the stability of geon configurations, because there exists no conceptual framework for analyzing this issue near the Planck scale and also at the quantum level. For scalar field wormholes, both charged and electrically neutral, the problem of stability was considered at the classical level~\cite{Gonzalez2009, Novikov2009, Bronnikov2012}, that is, first of all, for astrophysical scales. It was shown that these wormholes are unstable, with perturbations growing exponentially or linearly with proper time and occurring primarily near the throat. For topological geons with critical parameters, this instability implies that the growth of the perturbations seems very slow to a distant observer. Second, it is also possible that dark matter particles may interact with each other not only gravitationally. In particular, although dark matter is usually considered to be uncharged, it may consist of charged massive particles with ${q\sim10^{-13}\div10^{-6}e}$~\cite{Munoz2018, Meissner2019, Asjad2025}; this assumption gives us an additional parameter in order to explain current observations and operate more flexibly with theoretical models.

\vspace{2ex}\noindent
{\bf Conflicts of interest.} The author declares that he has no conflict of interest.

\vspace{2ex}\noindent
{\bf Acknowledgments.} The author thanks the participants of the MQFT-2025 conference for discussions on the talk, which resulted in this work.

\vspace{2ex}\noindent
{\bf Funding.} This work was funded from the budget of Tver State University.

\end{document}